\documentclass[journal]{IEEEtran}

\usepackage{xcolor}

\usepackage[T1]{fontenc}
\usepackage[latin9]{inputenc}
\usepackage{array}
\usepackage{units}
\usepackage{multirow}
\usepackage[pdftex]{graphicx} 
\usepackage{subcaption}
\usepackage[cmex10]{amsmath}
\usepackage{amssymb}
\usepackage{amsfonts}
\usepackage{amsmath}
\usepackage{supertabular}
\usepackage{times}
\usepackage{amsthm}
\usepackage{floatrow}
\usepackage{tikz}
\usepackage{float}
\usepackage{mathtools}
\usepackage{textcomp}
\usepackage{algpseudocode}
\usepackage[ruled,vlined]{algorithm2e}
\usepackage{adjustbox}
\usepackage[justification=centering, font=small]{caption} 
\usepackage{epstopdf}
\usepackage{tabularx}
\usepackage{nccmath}
\usepackage{mathtools}

\theoremstyle{plain}
\newtheorem{thm}{Theorem}

\newtheorem{lemma}{Lemma} 
\newtheorem{assumption}{Assumption}
\newtheorem{remark}{Remark}

\newfloatcommand{capbtabbox}{table}[][\FBwidth]

\newcommand{\vect}[1]{\boldsymbol{#1}}

\graphicspath{{./Figures/}}

\usepackage{cite}

\begin{document}

\title{Cooperative Deep $Q$-learning Framework for Environments Providing Image Feedback}

\author{Krishnan~Raghavan,~\IEEEmembership{Member,~IEEE,}
        Vignesh~Narayanan,~\IEEEmembership{Member,~IEEE,}
        and~Sarangapani~Jagannathan,~\IEEEmembership{Fellow,~IEEE}
\thanks{The project or effort undertaken is sponsored in part by the Department of the Navy, Office of Naval Research Grant N00014-21-1-2232,  Department of Army Cooperative Research Agreement W911NF2120260, and Intelligent Systems Center at Missouri University of Science and Technology, Rolla. Any opinions, findings, and conclusions or recommendations expressed in this material are those of the author(s) and do not necessarily reflect the views of the Office of Naval Research and Department of Army.}
\thanks{R. Krishnan is with the Mathematics and Computer Science Division, Argonne National Laboratory, USA; N. Vignesh is with the AIISC and the Department of Computer Science and Engineering, University of South Carolina, Columbia, USA, and S. Jagannathan is with the Department of Electrical and Computer Engineering, Missouri University of Science and Technology, Rolla, MO USA.}}

\maketitle
\begin{abstract}
In this paper, we address two key challenges in deep reinforcement learning setting, sample inefficiency and slow learning, with a dual NN-driven learning approach. In the proposed approach, we use two deep NNs with independent initialization to robustly approximate the action-value function in the presence of image inputs. In particular, we develop a temporal difference (TD) error-driven learning approach, where we introduce a set of linear transformations of the TD error to directly update the parameters of each layer in the deep NN. We demonstrate theoretically that the cost minimized by the error-driven learning (EDL) regime is an approximation of the empirical cost and the approximation error reduces as learning progresses, irrespective of the size of the network. Using simulation analysis, we show that the proposed methods enables faster learning and convergence and requires reduced buffer size (thereby increasing the sample efficiency).
\end{abstract}


\section{Introduction}\label{sec: introduction}
Control of complex systems is an important research thrust in the domain of science and engineering \cite{mahmud2018applications,luong2019applications,nguyen2020deep}.  In a typical application, sensors' measure physical quantities such as position, velocity, pressure, and temperature. These measurements are then utilized to design controllers that solve control/decision making tasks. However, in many applications such as the self-driving autonomous cars or control tasks performed by humans, sensory inputs are often encoded as images and/or audio signals. More recently, deep learning has enabled NNs to decode complex abstractions from measurement data. For instance, in the context of reinforcement learning (RL), the deep RL (DRL) facilitates complex decision-making problems involving high-dimensional state such as images and audio inputs and action spaces \cite{mnih2015human}. Despite decades of research, learning controllers for complex physical systems cannot be designed with performance guarantees when the sensors generate a sequence of images or audio signals~\cite{9284628} as feedback data for learning. One of the key reasons is that, deep neural networks (DNNs) with many hidden-layers  are required to extract complex abstractions from images or audio signals~\cite{MhaskarLP16}. However, the complex compositional structure of a deep NN does not allow for traditional Lyapunov theory-based analysis to provide any suitable conclusion about the performance of the DNN-based controller without making severe limiting assumptions.

A typical RL task is composed of three steps: (1.) an agent selects an action and utilizes it to actutate or effect change in the environment, (2.) the environment,  as a consequence of the action, transitions from its current state to a new state, and (3.) the environment then provides a  scalar reinforcement/reward signal as feedback. By performing these three steps repeatedly, the agent explores the environment and accumulates experiences or knowledge regarding the environment. Based on these experiences collected over time, the agent learns the best sequence of actions from any situation/state \cite{ref_extending_Iwata} to achieve a desired objective/goal. 

The desired objective is typically described as a function of the temporal difference (TD) error and this objective is estimated through an RL agent~($Q-$network parameterized by NN) and its clone~(a target)  \cite{mnih2015human}. A typical DRL learning scheme such as the one introduced in \cite{mnih2015human} aims to successively nudge the target network towards the optimal $Q-$function while the $Q$-network is trained to be close to the target. The key  distinction between the target in a DRL task with that of supervised learning task is that the targets in a DRL problem are non-stationary~\cite{mnih2015human}. Consequentially, efficient learning requires accurate approximation of the target, i.e., the optimal $Q$-function and efficient learning is pivotal on two key conditions: efficient use of experiences and sufficient exploration of the parameter or weight space.  

To enhance the process of collecting and efficiently utilizing experiences  within high-dimensional state (such as images and audio inputs) and action spaces \cite{ref_synth_Leonetti, ref_general_fact_domain_Hester, ref_general_human_robot_Hester, ref_model_visual_Massoud}, an experience replay strategy \cite{ref_self_improve_Lin, ref_batch_rein_learn_Kalyanakrishnan, ref_residential_demand_Ruelens, ref_Qlearn_exper_replay_Pieters} is usually employed in RL schemes. However, the memory space to store the experiences may be limited or expensive in many applications, and therefore, a mechanism to omit irrelevant experiences is desirable and have been considered~\cite{ref_backward_Qlearning_Wang, ref_Qlearn_exper_replay_Pieters, dist_prior}.  Despite appropriate sample selection strategies, useful experiences may be discarded. For instance, experiences collected initially, that are near the initial states, may be useful, but when some experiences are omitted based on time or reward, these experiences may be discarded. Even when these initial experiences are not discarded, they have the tendency to bias the learning procedure, \cite{thrun1993issues, mnih2015human} as the agent may spend more time exploring states dictated by the initial experiences but these experiences do not contribute positively to the overall objective of the agent. Despite several successful applications of DRL, these strategies of discarding or utilizing experiences, in general, are typically sample-inefficient \cite{schaul2015prioritized,jiang2020model}.

On the other hand, a tractable and efficient training strategy to use these experiences and update the network parameters is essential~\cite{ref_fmrq_multi_Zhang, ref_moo_mdp_Silva}. The most common strategy for training DNNs is the back-propagation/stochastic gradient descent (SGD) algorithm~\cite{mnih2015human}. Despite promising results, SGD suffers from issues such as vanishing gradient problem and slow learning~\cite{lecun2015deep,Ragh2017error,hardt2016train}. Many alternative approaches have been introduced in the literature to target these issues~\cite{Ragh2017error, lee2015difference, nokland2016direct, donoho2003hessian, reed2014training}. One prominent method is a direct error driven learning (EDL) approach that was proposed to train DNNs \cite{Ragh2017error} in the context of supervised learning. However, the EDL approach was not evaluated in DRL applications with image pixels as the NN inputs and a theoretical analysis of the EDL algorithm is a missing component in \cite{Ragh2017error}.


To address these issues, we introduce a dual-NN driven exploratory learning approach, wherein we improve sample efficiency using two NNs that are updated alternatively using the EDL update rule. We refer to this dual-NN driven alternative update mechanism as the \emph{cooperative update strategy} (coop).  In particular, the two DNNs estimate and minimize the desired objective~(a function of TD errors). Each DNN alternatively behaves as the target and the $Q-$network, therefore,  each NN is allowed to exploit, while behaving as target, and explore, while behaving as $Q-$network. 

In this setting, we adapt and expand the direct EDL rule proposed in \cite{Ragh2017error} to the case of DRL. In the EDL update rule, we transform the temporal difference (TD) error onto each layer in the deep NN through a non-singular matrix (linear transformation). Since the choice of the transformation matrix is critical, we construct this matrix for the purpose of increased exploration. We show that the solutions obtained with our scheme are as good as the typical SGD.  We show that our coop approach a) improves sample efficiency; b) enables better exploration; c) results in faster convergence and d) provides better performance in the context of online games using DRL. The contributions of this paper include: (1) alternative optimization-driven online learning approach for DRL involving high-dimensional state and action spaces to address sample efficiency; (2) extension of direct error-driven to RL with the cooperative learning strategy in order to mitigate vanishing gradient problem; (3) convergence analysis of the learning scheme. 
\section{Background}\label{sec: background}

In this section, we begin with a brief background on the RL problem, and then, introduce the notations used in the paper along with DNNs. Finally, we formally introduce the problem considered in this work. Additional background details are also provided in the Appendix.
\subsection{RL Preliminaries}
The RL problems can be formulated as Markov decision processes (MDPs) composed of a state-space $\mathcal{S}$, an action space $\mathcal{A}$, a state-transition function ${T}:\mathcal{S}\times\mathcal{A}\to\mathcal{S}$, and a reward function ${R}$ \cite{sutton2018reinforcement}. During the learning process, an RL agent perceives the environmental information, described by the state ${x}(k) \in \mathcal{S}$ and takes an action $u(k) \in  \mathcal{A}$ so that the environment transits from the current state to the next state $ {x}(k+1)$ based on the transition function ${T}$. The action $u(k)$ is, in general, based on a policy $\pi:\mathcal{S}\to \mathcal{A}$, and in this process, the agent receives an external scalar reward/reinforcement signal $r(k+1) \in  \mathcal{R}$ from the environment. The set composed of $\{ x(k), u(k),  x(k+1), r(k+1)\}$ is collected as experience by the RL agent to learn the value function or the $Q$-function. In this work, we utilize a DNN-based architecture to learn and approximate $Q$-functions in control/decision making problems concerned with systems/environments using data (from the environment) in the form of images.
\subsection{DNNs and notations}
Consider a feed-forward NN with $d$ layers denoted using a parametric map $y(\vect{z}; \vect{\theta})$, where $\vect{z}$ is the input and $\vect \theta$ denotes the weight parameters. In particular, let  $\vect{\theta} = [\vect{W}^{(1)} \cdots \vect{W}^{(d)}]$ denote the ideal weights of the DNN such that the parametric map $y(\vect{z}; \vect{\theta})$ is an approximation to a smooth nonlinear function $\mathcal F(\vect z)$ with compact support and an approximation error $\varepsilon$~\cite{paper3_approximation} such that $\mathcal F(\vect z) = y(\vect{z},\vect{\theta})+\varepsilon.$  The term $f^{(i)}$, for $i = 1, \ldots, d$ denotes the layer-wise activation functions (applied component-wise to the vector input) corresponding to the $d$ layers. In our proposed framework, we will employ such NNs to approximate the unknown function $\mathcal F(\vect{z})$. Let the estimated weights be denoted by $\hat{\vect{\theta}}  = [\hat{\vect{W}}^{(1)} \cdots \hat{\vect{W}}^{(d)}]$ such that the estimated net is denoted by $\hat{y}(\vect{z},\hat{\vect{\theta}})$, where the symbol $\hat{(\cdot)}$ is used to denote the estimated quantities. 
Additionally, throughout the paper, we use standard mathematical notations. Specifically, we use $\mathbb{N}$ and $\mathbb{R}$ to denote the set of natural numbers and real numbers, respectively. In the analysis presented in this paper, we use $\|.\|$ to denote the Euclidean norm for vectors and Frobenius norm for matrices. In the following, we formally introduce the problem considered in this paper.
\subsection{Problem description}
We consider an RL problem, where the dynamics or the transition function, $T$, of the environment are given as
    \begin{equation}
	    \begin{aligned}
    	    \vect{x}(k+1) = {T}(\vect{x}(k), \vect{u}(k), \vect{\omega}(k)),
    	 \end{aligned}
    	 \label{eq1}
    \end{equation}
where $\vect{x}(k) \sim p( \mathcal{S} )$ is the state of the system at the $k^{th}$ sampling instant, $\vect{u}(k)$ (actions) is the input to the system, and $\vect{\omega}(k)$ is some internal fluctuation, e.g., noise.  We will define a compact set $\mathcal{S}$ to describe all the states that can be observed by the agent and $p( \mathcal{S} )$ describes a distribution on the states. Typically, the exact dynamics of the system are unknown at any sampling instant $k$. We consider applications where the system output (i.e., the feedback data) is a sequence of images or snapshots. The input to the (learning) model are raw images that are pre-processed. We follow the pre-processing procedure detailed in \cite{mnih2015human} and the input to our network is four images of size $84 \times 84.$ One key difference between standard learning schemes and DRL is that the images are not the states of the system. These images encode the states and it is the job of the NN to extract the states.  Hence, the exact states are unknown to the model and must be extracted from the images. For instance, consider the game {\em breakout}, where the state of the game is represented by the position of the bar at the bottom of the screen. Since, the exact position of the bar is unknown, the NN must extract it from the images of the screen. With this feedback information, we seek to train an RL agent to learn a sequence of actions that yields maximum cumulative reward. We denote the reward functions by $r(\vect{x}(k), \vect{u}(k))$, which is a function of both the state and the action or control inputs such that for a given~(state, action pair) $(\vect{x}(k), \vect{u}(k))$ and an action policy $\pi$, the $Q$-function is given as
    \begin{equation}
    	\begin{aligned}
    	    Q^{\pi}(\vect{x}(k),\vect{u}(k)) =\sum_{k = 0}^{\infty} \gamma^{k} r(\vect{x}(k), \vect{u}(k)),
    	 \end{aligned}
    \end{equation}
where $\gamma \in (0,1)$ is the discount factor. In a typical DRL environment, the reward and images are obtained from an emulator. Our goal is to utilize the state and reward information from the environment to develop a learning framework to iteratively learn the optimal sequence of actions for a given task. Since, the optimality for a sequence of actions is determined by the $Q-$function, we seek to learn the optimal $Q-$function using NN, and since the feedback is in the form of images, we incorporate DNNs in our architecture.
\section{Proposed Method}\label{sec: methods}
In this section, we first introduce the optimization problem and the overall learning architecture. Later, we introduce the learning scheme to train the DNNs in the proposed architecture.
\subsection{RL Approximation of $Q$-values} 
The RL agent receives feedback information from the environment in the form of images, the RL agent must learn a map between the inputs and the $Q$ values corresponding to a fixed policy~(a greedy policy is considered in this paper). To update the $Q$ values, we use the Bellman's principle of optimality~\cite{sutton2018reinforcement} such that the optimal $Q$ function satisfies
    \begin{equation}
    	\begin{aligned}\label{eq:bellman}
    	   Q^{*}(\vect{x}(k), \vect{u}(k))  = max_{u(k)} (r(\vect{x}(k), \vect{u}(k))  \\ + \gamma^{k} (Q^{*}(\vect{x}(k+1), \vect{u}(k+1)) ),
    	 \end{aligned}
    \end{equation}
where $Q^*$ is the optimal $Q$-function. Let $\rho(\mathcal{A})$ describe a distribution over $\mathcal{A}$. The greedy policy or the optimal control is given as 
     \begin{equation}
    	\begin{aligned}
    	    \vect{u}^{*}(\vect{x}(k))  = argmax_{\vect{u}(k) \sim \rho} ( Q^{\rho}(\vect{x}(k), \vect{u}(k)) ),
    	 \end{aligned}
    \end{equation}
 where we use the notation $\rho$ to denote  $\rho(\mathcal{A})$. When the dynamics are known as in a linear system, the evaluation of the  optimal input to the system is well-known \cite{lewisoptimal}. However, in most general cases, the system/environment dynamics are often unknown. Therefore, the optimal $Q$-function, and in turn the control policy, have to be approximated.  In this work, we will approximate the $Q$-function value for each action in the set $\mathcal{A}$ by defining a generic parametric map, $y(\vect{x}(k), \vect{u}(k) ; \vect{\theta})$ with an ideal set of parameters $\vect{\theta}$ such that 
    \begin{equation}
    	\begin{aligned}
    	    \vect{u}^{*}(\vect{x}(k))  = argmax_{ \vect{u}(k) \in \mathcal{A}} (y(\vect{x}(k), \vect{u}(k) ; \vect{\theta})).
    	 \end{aligned}
    \end{equation}
The learning problem is to find an approximation $\hat{y}$ with parameters $\hat{\vect{\theta}}(k)$ for the function $y$. We employ DNNs as function approximators to learn $\hat{y}$. To train the weights of the DNN, the cost is defined as the expected value over all the states and all the actions such that $ \mathbb{E}_{\vect{x}(k) \sim p(\mathcal{S}), \vect{u}(k) \sim \rho(\mathcal{A}) }[ J(\hat{\vect{\theta}}(k) )],$ where
\begin{equation}
    	\begin{aligned}
    & J(\hat{\vect{\theta}}(k) ) = \frac{1}{2}(y( \vect{x}(k), \vect{u}(k) ; \vect{\theta}(k) ) - \hat{y}( \vect{x}(k), \vect{u}(k); \hat{\vect{\theta}}(k)) )^{2},
    	 \end{aligned}
\end{equation}
with target $y( \vect{x}(k), \vect{u}(k); \vect{\theta}(k) )   = r(\vect{x}(k), \vect{u}(k))  + \gamma^{k}{\hat{Q}^{*}(\vect{x}(k+1), \vect{u}(k+1))}$, coming from the Bellman equation with $\hat{Q}^{*}$ denoting the approximation of the optimal $Q$-value obtained by extrapolating the current $Q$-function estimate. Therefore, we may write
    \begin{equation}
    	\begin{aligned}
    	   &\hat{\vect{\theta}}(k) = arg~min_{\hat{\vect{\theta}}(k)} \mathbb{E}_{\vect{x}(k) \in \mathcal{S}, \vect{u}(k) \sim \rho(\mathcal{A}) }&  \\ &\big[ J_E(\vect{x}(k), \vect{u}(k), \hat{y}( \vect{x}(k), \vect{u}(k); \hat{\vect{\theta}}(k)) )\big], & 
    	\end{aligned}
    	 \label{eq_main_opt}
    \end{equation}
    with
        \begin{equation}
    \begin{aligned}
 J_E(\vect{x}(k), \vect{u}(k) , \hat{y}( \vect{x}(k), \vect{u}(k); \hat{\vect{\theta}}(k)) )&=& \nonumber \\  \frac{1}{2} \big[r_{k} + \gamma^{k}\hat{Q}_{k+1}^{*} - \hat{y}_{k} \big]^2 &=& \frac{1}{2} \big[ \epsilon_{k} \big]^2, \nonumber \\
    	 \end{aligned}
    	 \label{eq_main_opt}
    \end{equation}
 To simplify notations, we have used $r_{k}$ to denote $r(\vect{x}(k), \vect{u}(k))$  and denote {$\hat{Q}^{*}(\vect{x}(k+1), \vect{u}(k+1))$} by $\hat{Q}_{k+1}^{*}$ and $\hat{y}(k; \hat{\vect{\theta}}(k))$ by $\hat{y}_{k}$. Note that $r_{k} + \gamma^{k}\hat{Q}_{k+1}^{*} - \hat{y}_{k}$ is the temporal difference (TD) error denoted as $\epsilon_{k}$ from here on. The empirical cost which we seek to minimize is the squared temporal difference error.
\subsection{Deep Q Learning~\label{sec:dql}}
 \begin{figure}[h]
\includegraphics[scale=0.8, width = \columnwidth, keepaspectratio = True]{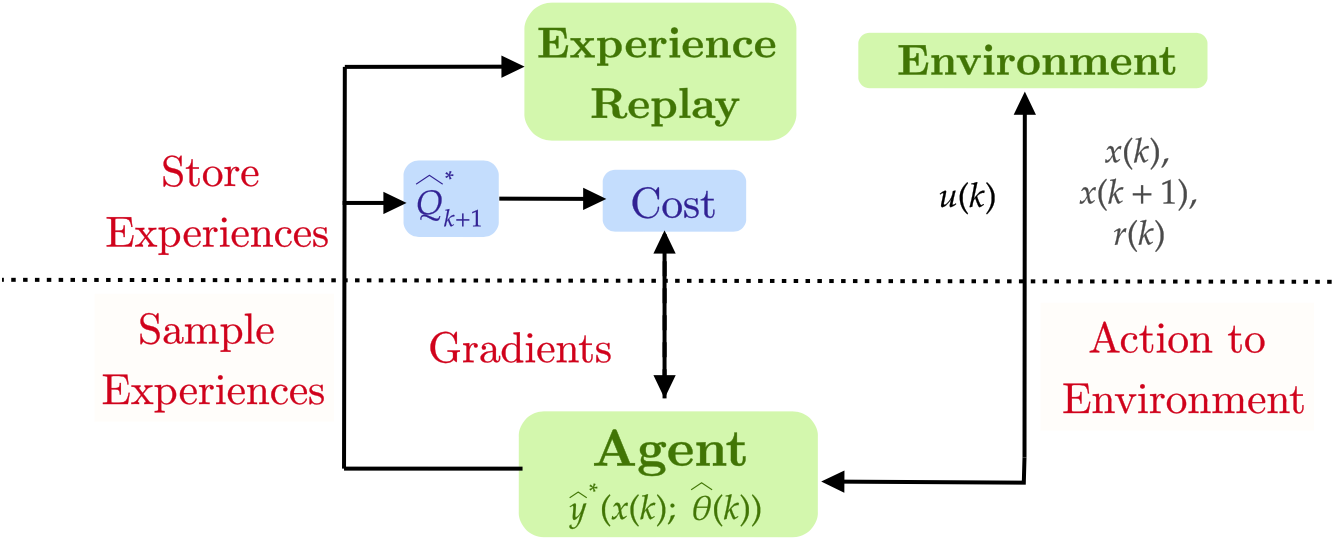}
    \caption{DQN Algorithm }
    \label{fig:DQNAlgo}
\end{figure} A learning algorithm to solve the optimization problem in  Eq. \eqref{eq_main_opt} was provided in \cite{mnih2015human}. In \cite{mnih2015human}, the total learning process was split into episodes. For each episode, there are several plays, each play is comprised of five basic steps~(refer Fig. \ref{fig:DQNAlgo}): (1) current state is inferred from images provided by the environment; (2) this state information is utilized by the $Q$-network to generate actions; (3) the actions are then presented to the environment; (4) the environment evolves and provides the rewards and the next state via images; (5) the tuple of (current state, next state, action, reward) is stored into an experience replay array as batches; (6) Finally, batches are sampled from the experience replay array to update the network parameters using the gradient of $J_E(\vect{x}(k), \vect{u}(k), \hat{y}( \vect{x}(k); \hat{\vect{\theta}}(k))) $.

In a typical DRL algorithm ~\cite{mnih2015human}, the TD error~$(\epsilon_k = r_{k} + \gamma^{k}\hat{Q}_{k+1}^{*} - \hat{y}_{k})$ is computed by extrapolating the current $Q$-network to 'guess' the $Q$-values of future states~(the target). This is typically achieved by defining two networks~($Q-$network and target) and periodically copying the weights of the  current $Q$-network to the target. Then, the current state is used as input to the $Q$-network to generate $\hat{y}_{k}$ while the future state is used as an input to the target to generate $\hat{Q}_{k+1}^{*}.$ Using the $\epsilon_k$ we update the weights of $\hat{y}( \vect{x}(k); \hat{\vect{\theta}}(k))$ during the next $C$ plays.

\subsection{Cooperative dual network architecture~(coop)}
\begin{figure}[h]
    \centering
    \includegraphics[width=\columnwidth,keepaspectratio]{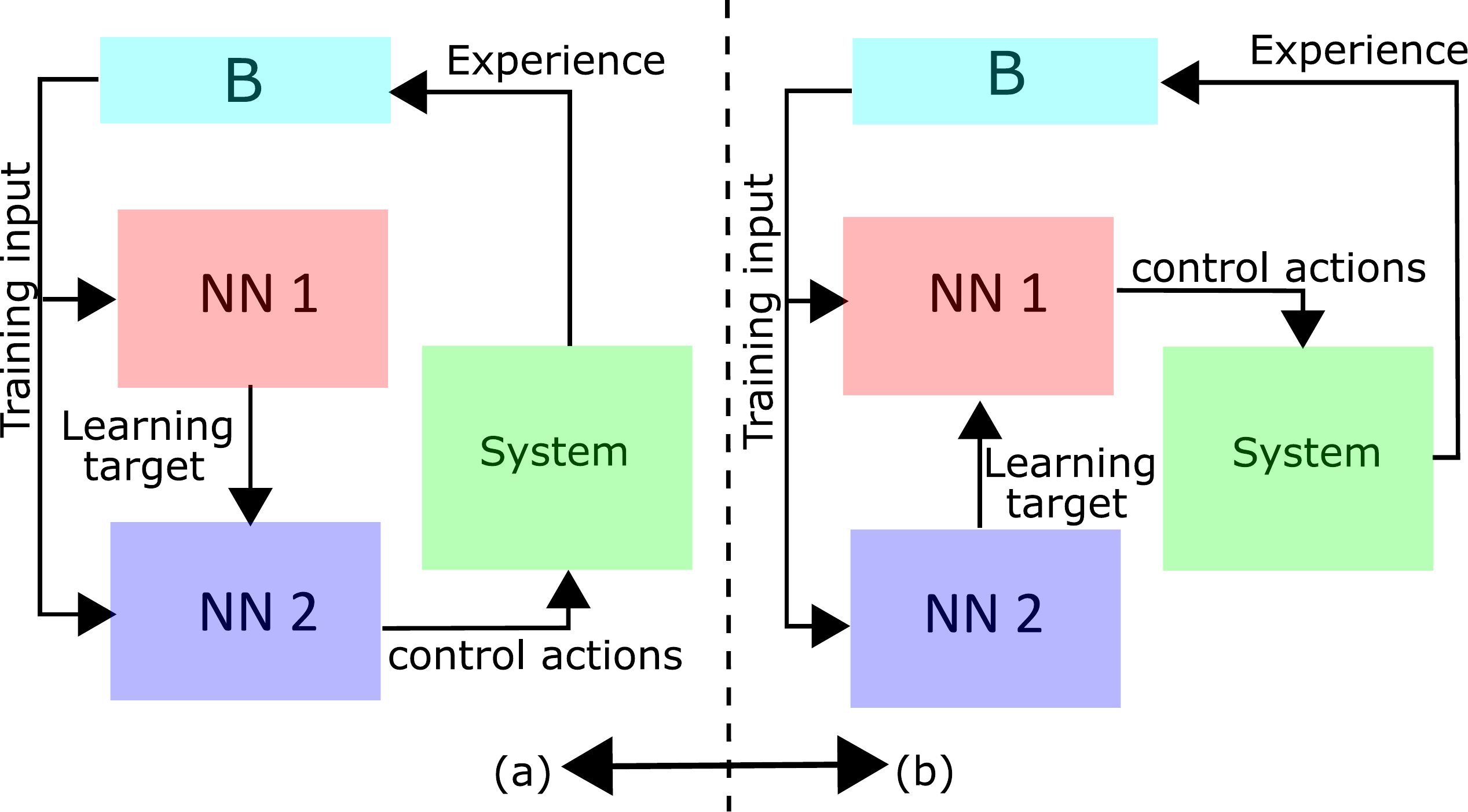}
    \caption{Schematic diagram illustrating the Co-op dual DQN architecture. During learning the NNs~(NN 1 and NN 2) switch roles as shown in (a) and (b). In (a) NN 2 is trained using the experience stored in the memory buffer with targets given by NN 1; and in (b) the NNs swap their roles.}
    \label{fig:schematic}
\end{figure}
Our coop strategy utilizes the algorithm from \cite{mnih2015human}~(also described in \ref{sec:dql}) but incorporates two DNNs initialized at random with different hyper-parameters-(weights and biases) instead of the periodically cloning the $Q-$network every C-plays. 

 Our coop strategy is comprised of two steps: first, we designate one of the NNs, say NN 1, to generate the targets~(future state is given as input to NN 1) and NN 2 for generating action~(current state is given as input), where we update the weights of NN 2. Subsequently, in the following iteration, NN 2 generates the target and  NN 1 generate the action. We always update the network that generates the action and with alternative updates, the two NNs play the role of learning the $Q$ function and the target respectively~(see Fig. \ref{fig:schematic}). 

The rationale behind the proposed learning scheme where the NNs have cooperative functionality are two fold. We expect that the NNs encode the training data from the samples much more efficiently while reducing the memory footprint and necessity of raw data stored in the memory buffer, thereby improving sampling efficiency. Secondly, since each NN obtains target from the other NN during training, these NNs are expected to converge asymptotically while improving the robustness of the learning. In the numerical experiments presented in Section \ref{sec: simulations}, we show that equipped with a exploratory error driven learning rule, the proposed learning architecture is robust, improves the sample utilization and performs better with lesser buffer memory, and converges faster when compared with the existing results \cite{mnih2015human}.

\subsection{Error driven TD learning}
To minimize the empirical cost, gradient-based update rules are commonly employed \cite{lewis1998neural, rumelhart1988learning}. These update rules are typically written as
    \begin{equation}
            \label{eq8}
            \begin{aligned}
            	\hat{\vect{W}}^{(i)} (k+1) = \hat{\vect{W}}^{(i)} (k) + \alpha \vect{\Delta}^{(i)}(k), 
            \end{aligned}
    \end{equation}
where $\alpha >0$ is the learning rate, $i$ is the layer number, $k$ is the sampling instant, and $\vect{\Delta}^{(i)}(k)$ is the parameter adjustment made along a descent direction by using the gradient of the error propagated backwards from the output layer of the network \cite{hardt2016train,lewis1998neural}. To simplify notations, we switch to a subscript notation instead of expressing the sampling instance in parenthesis, so that \eqref{eq8} is rewritten as
	$\hat{\vect{W}}^{(i)}_{k+1} = \hat{\vect{W}}^{(i)}_{k} + \alpha \vect{\Delta}^{(i)}_{k}$. 
From here on, for brevity, the notation of the expected value operator is suppressed, and we refer to the expected value of the cost as just the cost value. Furthermore, we add a regularization term to the cost function such that the revised cost is written as
\begin{align}
    {{H}(\hat{\vect{\theta}}_k)}= \big[ J_{E}(\hat{\vect{\theta}}_{k})  +   \sum_{i = 1}^{d} \lambda^{(i)} R_{k}(\hat{\vect{W}}^{(i)}_{k})\big],
    \label{eq:orignal}
\end{align}
where $\lambda^{(i)} > 0$ is the decay coefficient, $R_{k}(\hat{\vect{W}}^{(i)}_{k})$ denotes the function of regularization applied on the weights $\hat{\vect{W}}^{(i)}_{k}$, and the optimal cost~${H}^{*}(\hat{\vect{\theta}}_k)$ satisfies
\begin{equation}
        \begin{aligned}
	        {H}^{*}(\hat{\vect{\theta}}_k) =  \underset{\hat{\vect{\theta}} \in \vect{\Omega} }{\text{min}} \quad \big[{J_{E}(\hat{\vect{\theta}}_{k})  +  \sum_{i = 1}^{d} \lambda^{(i)}  R_{k}(\hat{\vect{W}}^{(i)}_{k})}\big],
        \end{aligned}
    \label{eq9}
\end{equation}
where $J_E(\hat{\vect{\theta}}_{k})$ is used in place of $J_E(\vect{x}(k), \vect{u}(k); \hat{\vect{\theta}}(k)).$
The weight adjustment at each iteration $k$ is given as 
\begin{IEEEeqnarray}{rCl}
    \vect{\Delta}^{(i)}_{k}  &=&  -\big[ \nabla_{\hat{\vect{W}}^{(i)}_{k}}   H(\hat{\vect{\theta}}_{k})  \big],\nonumber \\  &=& -\big[ {\vect{\delta}}^{(i)}_{k}  + \lambda^{(i)}  \nabla_{\hat{\vect{W}}^{(i)}_{k}} R(\hat{\vect{W}}^{(i)}_{k}) \big],
    \label{eq11}
\end{IEEEeqnarray}   
where the term $\nabla_{\hat{\vect{W}}^{(i)}_k}(.)$ denotes the gradient of (.) with respect to the NN weight  $\hat{\vect{W}}^{(i)}_{k}$, the second term in this update rule depends on the choice of $R(\hat{\vect{W}}^{(i)}_{k})$, and $\vect{\delta}^{(i)}_{k}$  is obtained by applying the chain rule to compute the gradient of the cost with respect to weight.
Specifically, applying the chain rule to compute layer-wise gradient through error backpropagation \cite{Ragh2017error,lecun2015deep}, a generalized expression for ${\vect{\delta}}^{(i)}_{k} $ can be derived as
    \begin{equation}
    \label{eq2_Grad}
	\begin{aligned}
      {\vect{\delta}}^{(i)}_{k}  = f^{(i-1)}(\vect{x}) \big[ \prod_{j = d}^{i+1} diag(\nabla f^{(j)}(\vect{x})) \hat{\vect{W}}^{(j)} \big]  \\ diag(\nabla f^{(i)}(\vect{x}))\epsilon_{k} \vect{I}^{(i)}.
      \end{aligned} 
      \end{equation}
We now denote $ \prod_{j = d}^{i+1}( diag(\nabla f^{(j)}(\vect{x})) \hat{\vect{W}}^{(j)}) diag(\nabla f^{(i)}(\vect{x})) $ as $\vect{\mathcal{T}}^{(i)}(\vect{x})$ and simplify Eq. \eqref{eq2_Grad} to get 
\begin{equation}
\label{eq12}
\begin{aligned}
    {\vect{\delta}}_{k}^{(i)}   =  f^{(i-1)}(\vect{x}) \epsilon_{k} \vect{\mathcal{T}}^{(i)} \vect{I}^{(i)}, 
\end{aligned}
\end{equation}
where $\epsilon_{k} = y_k-\hat{y}_k=r_{k} + \gamma^{k}{\hat{Q}_{k+1}^{*}} - \hat{y}_{k}$ is the temporal difference error~(TD error). Observe that the temporal difference error $\epsilon_{k}$ propagates through transformation $\vect{\mathcal{T}}^{(i)}$ to influence learning. Moreover, for a fixed $\vect{x},$ $\vect{\mathcal{T}}^{(i)}(\vect{x})$ is a linear approximation of the gradients in the neighborhood of the weights. Since, this approximation of the gradient is a matrix, the singular vectors of $\vect{\mathcal{T}}^{(i)}$ dictate the directions of learning. Their magnitude is described by the singular values of $\vect{\mathcal{T}}^{(i)}$. Each diagonal element in $\vect{\mathcal{T}}^{(i)}$ is the product of the derivative of the layer-wise activation functions. In other words, the singular values of $\vect{\mathcal{T}}^{(i)}$ would also tend towards zero as the number of layers in the DNN increase, otherwise known as the vanishing gradients issue~(see \cite{pascanu2013difficulty} and the references therein for details).

Therefore, to address this issue, we introduce a user-defined feedback matrix $\vect{B}^{(i)}_{k}$ such that $\vect{B}^{(i)}_{k} (\vect{B}^{(i)}_{k})^{T}$ is positive definite, to take the role of $\vect{\mathcal{T}}^{(i)}_{k}$ in \eqref{eq12}. The new feedback, denoted as $\vect{\sigma}_{k}^{(i)}$, is then given as
\begin{equation}
	\label{eq:edl_update}
	\begin{aligned}
    \vect{\sigma}_{k}^{(i)}  = [f^{(i-1)}(\vect{x})]  \epsilon_k \vect{B}^{(i)}_{k}.
 	\end{aligned}
	\end{equation}
With this definition of feedback, the new layer-wise cost  function denoted with $\mathcal{H}^{(i)}(\vect{x}(k), \vect{u}(k); \hat{\vect{W}}^{(i)})$ is defined to be
 \begin{IEEEeqnarray}{rCl}
    \mathcal{H}^{(i)}(\vect{x}(k), \vect{u}(k); \hat{\vect{W}}^{(i)}) &= \frac{1}{2} \big[tr( ( \vect{\sigma}_{k}^{(i)} )^{T} \vect{P} {\hat{\vect{W}}^{(i)}}_{k} )&\nonumber \\ &+ \lambda^{(i)} R({\hat{\vect{W}}^{(i)}_{k}}) \big],& \label{eq:eq_mod_cost}
\end{IEEEeqnarray} 
where $tr(.)$ is the trace operator and $\vect{P}$ is  a positive definite symmetric matrix of choice which replaces the identity matrix in the update for the gradient descent rule. Note that for each layer $i$, $\vect{\sigma}^{(i)}_{k}$ can be understood as the feedback provided by the overall cost $J_{E}(\hat{\vect{\theta}})$ towards controlling the layer $i$ of the DNN. \emph{Therefore, the layer-wise cost can be interpreted as the minimization of the correlation between $\vect{\sigma}^{(i)}_{k}$ and  $\hat{\vect{W}}^{(i)}_{k}$ under the constraint that $\|\hat{\vect{W}}^{(i)}_{k}\|$ is bounded (due to the regularization).} Finally, the overall cost denoted as $ \mathcal{H}(\vect{x}(k), \vect{u}(k); \hat{\vect{\theta}}(k))$ may be written as the sum of layer-wise costs as 
  \begin{equation}
    \begin{aligned}
        \mathcal{H}(\vect{x}(k), \vect{u}(k); \hat{\vect{\theta}}(k)) &=& \sum_{i=1}^{d}  \mathcal{H}^{(i)}(\vect{x}(k), \vect{u}(k); \hat{\vect{W}}_{k}^{(i)}),
     \end{aligned}
     \label{eq:eq_mod_cost_total}
\end{equation}
With the cost $ \mathcal{H}(\hat{\vect{\theta}}_{k})$ defined in Eq. \eqref{eq15}, the optimization problem for the direct error-driven learning can be rewritten as
\begin{equation}
  \begin{aligned}
	   &\hat{\vect{\theta}}(k) = arg~min_{\hat{\vect{\theta}}(k)} \mathbb{E}_{\vect{x}(k) \in \mathcal{S}, \vect{u}(k) \sim \rho(\mathcal{A}) }\big[ \mathcal{H}(\vect{x}(k), \vect{u}(k); \hat{\vect{\theta}}(k)) \big].
  \end{aligned}
\label{eq_modified_opt}
\end{equation}
The weight updates for each layer are defined as $\vect{\Delta}^{(i)}_{k} = - \nabla_{\hat{\vect{W}}^{(i)}_{k}} \mathcal{H}(\vect{x}(k), \vect{u}(k); \hat{\vect{\theta}}(k))$, which results in the update rule as defined in Eq. \eqref{eq11} with $\vect{\delta}^{(i)}_{k}$ replaced by $\vect{\sigma}_k^{(i)}$, and defined as in Eq. \eqref{eq:edl_update}. For simplicity, from here on, we will denote $ \mathcal{H}(\vect{x}(k), \vect{u}(k); \hat{\vect{\theta}}(k))$ as  $\mathcal{H}( \hat{\vect{\theta}}(k))$ and  $\mathcal{H}^{(i)}(\vect{x}(k), \vect{u}(k);\hat{\vect{W}}_{k}^{(i)})$ as $\mathcal{H}^{(i)}(\hat{\vect{W}}_{k}^{(i)})$.
 \begin{remark}
     The direct error driven learning (EDL)-based weight updates proposed in \cite{Ragh2017error} can be interpreted as an exploratory update rule, as the descent direction in the weight updates are assigned through $B^{(i)}_k$ matrix. To define the matrix $B^{(i)}_k$, we decompose it as $\vect{B}^{(i)}_{k}. = \vect{U}^{(i)}_{k}(\Sigma^{(i)}_{k}+s \times I^{(i)})\vect{V}^{(i)}_{k},$ where $\vect{U}^{(i)}_{k} \Sigma^{(i)}_{k} \vect{V}^{(i)}_{k}$ is the SVD of $\vect{\mathcal{T}}^{(i)}_{k}$ for a fixed $\vect{x}$ with $I^{(i)}$ being an identity matrix of appropriate dimensions and $s$ is a chosen perturbation.
  \end{remark}
  In the following, we analyze the EDL scheme and extract some insights on the role of the matrix $B^{(i)}_k$ introduced in the update rule to train NNs of $d$ layers.
\subsection{Analysis of EDL for training DNN}\label{sec: analysis}
In this paper, we approximate the optimization problem associated with \eqref{eq9} using \eqref{eq_modified_opt} and introduce the idea of exploration in the learning approach through the choice of perturbations in the design matrix $\vect{B}^{(i)}$. In this section, we seek to analyze different components of this learning methodology. In order to proceed further, we make the following assumptions. 
\begin{assumption}
    The objective function is a smooth function in its domain, and in particular, $J_{E} \in C^1$. In addition, for all $\vect{x}\in \mathcal{S}, \vect{u}\in \mathcal{A}$, and $\hat{\vect{\theta}} \in \Omega$, the empirical cost and its gradient are bounded, i.e., $J_{E}(.,.,\hat{\vect{\theta}}) \leq L$  and $\frac{\partial J_E}{\partial \hat{\vect{\theta}}}\le M$. There exists a positive constant $\vect{\theta}_B$ such that for any $ \hat{\vect{\theta}}, \vect{\theta} \in \Omega$, we have $max( \|\hat{\vect{\theta}}_k\|,  \|\vect{\theta}\|) \leq \vect{\theta}_B$. The activation function $f$ of the NNs are chosen such that for any $x\in \mathbb{R}$, $\|f(x)\|\le1$. Finally, $s \sim \mathcal{N}(0,1)$, where $\mathcal{N}$ denotes the normal distribution.
\label{ass1}
\end{assumption}

\begin{lemma}
    Let Assumption 1 be true. Consider the empirical cost with regularization given as $${{H}(\hat{\vect{\theta}}_k)}= \big[ J_{E}(\hat{\vect{\theta}}_{k})  + \lambda^{(i)}  \sum_{i = 1}^{d} R_{k}(\hat{\vect{W}}^{(i)}_{k})\big].$$ Then, the value of  ${{H}(\hat{\vect{\theta}}_k)}$ can be rewritten as $$ {{H}(\hat{\vect{\theta}}_k)} = \sum_{i=1}^{d} \frac{1}{2} \big[tr( ( \vect{\delta}_{k}^{(i)} )^{T}{\hat{\vect{W}}^{(i)}}_{k} ) + \lambda^{(i)} R({\hat{\vect{W}}^{(i)}_{k}}) \big] + \xi,$$ where $\xi =   J_{E}(\vect{\theta}) -\bigg(\frac{\partial J_{E}( \hat{\vect{\theta}}_k ) }{\partial \hat{\vect{\theta}}_k}\bigg)^T {\vect{\theta}}   - H.O.T$, where $H.O.T$ represents the higher order terms of ($\hat{\vect{\theta}}_k-\vect{\theta}$).
\end{lemma}
\begin{proof}
    The proof follows the standard linear approximation using Taylor series expansion of the function $J_E({\vect{\theta}})$. 
To see this, expand $J_{E}({\vect{\theta}})$ around $\hat{\vect{\theta}_k}$ to get 
\begin{align}
        J_{E}({\vect{\theta}})= \bigg[ J_{E}(\hat{\vect{\theta}}_k) + \bigg(\frac{\partial J_{E}( \hat{\vect{\theta} }) }{\partial \hat{\vect{\theta}}}\bigg)^T [{\vect{\theta}} -\hat{ \vect{\theta}}_k]+H.O.T \bigg].
\end{align}
Rearranging this equation, we get 
\begin{align}
         J_{E}(\hat{\vect{\theta}}_k) = \bigg[J_{E}({\vect{\theta}}) + \bigg(\frac{\partial J_{E}( \hat{\vect{\theta}_k }) }{\partial \hat{\vect{\theta}}_k}\bigg)^T [\hat{ \vect{\theta}}_k-{\vect{\theta}}]-H.O.T \bigg].
\end{align}
Define $ \xi =   J_{E}(\vect{\theta})-\bigg( \frac{\partial J_{E}( \hat{\vect{\theta}}_k ) }{\partial \hat{\vect{\theta}}_k } \bigg)^T {\vect{\theta}}  - H.O.T$ and write 
\begin{align}
        J_{E}(\hat{\vect{\theta}}_{k})=  \bigg( \frac{\partial J_{E}( \hat{\vect{\theta}}_k ) }{\partial \hat{\vect{\theta}}_k } \bigg)^T \hat{\vect{\theta}}_k + \xi
\end{align}
Note that $\frac{\partial J_{E}( \hat{\vect{\theta}}_k ) }{\partial \hat{\vect{\theta}}_k }  = \frac{\partial J_{E}( \hat{\vect{\theta}}_k ) }{\partial \hat{y} } \frac{\partial \hat{y} }{\partial \hat{\vect{\theta}}_{k}}.$
If we let $\frac{\partial \hat{y} }{\partial \hat{\vect{\theta}}_{k}} = [\frac{ \partial \hat{y}  }{\partial \hat{\vect{W}}^{(1)}_{k} }  , \frac{ \partial \hat{y} }{\partial \hat{\vect{W}}^{(2)}_{k} } , \cdots, \frac{ \partial \hat{y} }{\partial \hat{\vect{W}}^{(d)}_{k} } ]$ and  $\frac{\partial J_{E}( \hat{\vect{\theta}}_k ) }{\partial \hat{y} }  = \epsilon_{k}$, 
we get 
\begin{align}
      {{H}(\hat{\vect{\theta}}_k)} = \bigg[ \frac{ \partial \hat{y}  }{\partial \hat{\vect{W}}^{(1)}_{k} }  , \frac{ \partial \hat{y} }{\partial \hat{\vect{W}}^{(2)}_{k} } , \cdots, \frac{ \partial \hat{y} }{\partial \hat{\vect{W}}^{(d)}_{k} }  \bigg]^T \epsilon_{k} \nonumber \\  \bigg[ \hat{\vect{W}}^{(2)}_{k}, \hat{\vect{W}}^{(2)}_{k} , \cdots \hat{\vect{W}}^{(d)}_{k} \bigg] + \xi + \lambda^{(i)}  \sum_{i = 1}^{d} R_{k}(\hat{\vect{W}}^{(i)}_{k})\big],
\end{align}
which can be rewritten as 
\begin{align}
     {{H}(\hat{\vect{\theta}}_k)} = \sum_{i=1}^{d} \frac{1}{2} \big[tr( ( \vect{\delta}_{k}^{(i)} )^{T}{\hat{\vect{W}}^{(i)}}_{k} ) + \lambda^{(i)} R({\hat{\vect{W}}^{(i)}_{k}}) \big]+ \xi.
\end{align}
\end{proof}

\begin{thm}
Let Assumption 1 and Lemma 1 hold true. Consider the original optimization problem defined in Eq. \eqref{eq:orignal} with the cost denoted as $ H( \hat{\vect{\theta}}_{k} )$  and consider the modified optimization problem defined in Eq. \eqref{eq:eq_mod_cost_total} with the cost denoted as $\mathcal{H}( \hat{\vect{\theta}}_{k})$. Let the feedback matrix be defined as $\vect{B}^{(i)}_{k} = \vect{U}^{(i)}_{k}(\Sigma^{(i)}_{k}+s \times I^{(i)})\vect{V}^{(i)}_{k},$ where $\vect{U}^{(i)}_{k} \Sigma^{(i)}_{k} \vect{V}^{(i)}_{k}$ is the SVD of $\vect{\mathcal{T}}^{(i)}_{k}$ with $I^{(i)}$ being an identity matrix of appropriate dimensions. Let $s \sim \mathcal{N}(0,1)$ be a chosen perturbation, where $\mathcal{N}$ describes the normal distribution. Consider the conditions for $i=1,\ldots,d$ that $\| {\hat{\vect{W}}^i}_{k}\| \leq W_B$ and $\|f^{(i-1)}(\vect{x})\| <\sqrt{\eta},$  where $\eta$ is the total number of neurons at each layer. Choose $\vect{P}$ such that $\|\vect{P}\| \leq 1$. The square difference between  original cost~(Eq. \eqref{eq:orignal}) and the new cost~(Eq. \eqref{eq:eq_mod_cost_total}) is then given as
\begin{IEEEeqnarray}{rCl}
  \| H( \hat{\vect{\theta}}_{k} )-\mathcal{H}( \hat{\vect{\theta}}_{k} )  \| \leq  d \frac{|s|}{2} \|\epsilon_{k}\| W_B \sqrt{\eta} + \|\xi\|
\end{IEEEeqnarray} 
and $E\| H( \hat{\vect{\theta}}_{k} )-\mathcal{H}( \hat{\vect{\theta}}_{k} ) \| \leq \|\xi\|.$
\end{thm}
\begin{proof}
See \cite{Ragh_acc_arxiv}
\end{proof}

In the above theorem, we have shown that the direct error driven learning problem is an approximation to the original optimization problem with residuals controlled by the choice of the perturbations. In other words, the residual will be zero when the choice of perturbations are uniformly zero. 
\begin{thm}
Let Assumption 1 be true. Let the gradient of the empirical cost be given as $\frac{\partial J_{E}( \hat{\vect{\theta}}_{k} ) }{\partial \hat{\vect{\theta}}_{k}} = [ f^{(1)}(\vect{x}) \epsilon_{k} \vect{\mathcal{T}}^{(1)} \cdots f^{(d)}(\vect{x}) \epsilon_{k}  \vect{\mathcal{T}}^{(d)}]$. Consider the weight update rule as 
$\hat{\vect{\theta}}_{k+1} = \hat{\vect{\theta}}_{k} +  \vect{\Delta}_{k},  \vect{\Delta}_{k} = -\alpha(k) \times \frac{ \partial \mathcal{H}_{k}( \hat{\vect{\theta}}(k))  }{\partial \hat{\vect{\theta}}_{k} },$
with $\alpha(k) > 0$ and  $\frac{\partial \mathcal{H}_{k}( \hat{\vect{\theta}}_{k}) }{\partial \hat{\vect{\theta}}_{k}}  =  [ f^{(1)}(\vect{x}) \epsilon_{k}  \vect{{B}}^{(1)}+ \frac{\partial \lambda^{(i)} R(\hat{\vect{W}}^{(1)})}{\partial\hat{\vect{W}}^{(1)}}  \cdots f^{(d)}(\vect{x}) \epsilon_{k} \vect{{B}}^{(d)} + \frac{\partial \lambda^{(i)} R(\hat{\vect{W}}^{(d)})}{\partial\hat{\vect{W}}^{(d)}} ]$ with the feedback matrix defined as $\vect{B}^{(i)}_{k} = \vect{U}^{(i)}_{k}(\Sigma^{(i)}_{k}+s \times I^{(i)})\vect{V}^{(i)}_{k},$ where $\vect{U}^{(i)}_{k} \Sigma^{(i)}_{k} \vect{V}^{(i)}_{k}$ is the SVD of $\vect{\mathcal{T}}^{(i)}_{k}(\vect{x})$ for a given batch of data $\vect{x}$ with $I^{(i)}$ being an identity matrix of appropriate dimensions and $s$ is a known perturbation. Let $\vect{B}^{(i)}, i = 1, \cdots d$ be chosen with non zero singular values 
then $J_E$ converges asymptotically in the mean.
 \end{thm}
 \begin{proof}
See \cite{Ragh_acc_arxiv}
 \end{proof}
 \begin{remark}
     The EDL learning rule was introduced in \cite{Ragh2017error}, and  it was shown to perform very well in learning applications. In this paper, we attempt to uncover the reasons for the performance of the EDL. In this context, the analysis presented in this section reveals the following. The results of Lemma~1 and Theorem~1 reveal that the cost function which is minimized by the EDL is a first-order approximation of the empirical cost, and that the difference between these costs will reduce as the weights of the NN converge. Secondly, from Theorem 1, we show that the parameters of the EDL updates rule can be chosen such the the original cost function asymptotically converges, so that the training error~(temporal difference error) goes to zero asymptotically.   
 \end{remark}
 Equipped with the parameter update, we sketch our learning algorithm as follows. 
 
 \subsection{Algorithm}
 Our algorithm is summarized in Algorithm \ref{alg1}.  The training process is performed for a total of $M$ episodes. At the start of episode $1$, we initialize the two NNs that is $NN_1$ and $NN_2$. In each episode there are $K$ plays. In each play,  we provide an action to the environment and we receive a tuple comprised of (reward, state, next-state). This tuple and the action is stored in the experience replay buffer (denoted as $\mathcal{B}$ with buffer size $N$).  
 
\begin{algorithm}
	Initialize $\hat{\vect{\theta}}_{1}$~($Q_1$), $\hat{\vect{\theta}}_{2}$~$Q_2$, and buffer~$\mathcal{B}$ of capacity N\\
	\For{$episodes=1,2,3,...$}{
	    Set flag = True \\
		\For{$k = 1, 2, ... $}{
		    Observe image $x_{k},$ and  preprocess it\\
		    \eIf{flag == True}{
    		    With probability $\epsilon$ select a random action $a_{k}$;
    		    otherwise $a_{k} = argmax(Q_{1}(x_{k}))$;\\
    		    Execute $a_{k}$ in emulator and observe reward $r_{k}$ and image $x_{k+1}.$\\
    		    Store the tuple $(x_{k}, a_{k}, x_{k+1}, r_{k})$ into $\mathcal{B}$ after pre-processesing.\\
    		    Sample batch of tuples from $\mathcal{B}.$\\
    		    Calculate the error $r+\gamma Q_{2} - Q_{1}$ and update $\hat{\vect{\theta}}_{1}.$\\}{With probability $\epsilon$ select a random action $a_{k}$
    		    otherwise $a_{k} = argmax(Q_{2}(x_{k})).$\\
    		    Execute $a_{k}$ in emulator and observe reward $r_{k}$ and image $x_{k+1}.$\\
    		    Store the tuple $(x_{k}, a_{k}, x_{k+1}, r_{k})$ into $\mathcal{B}$ after pre-processesing.\\
    		    Sample batch of tuples from $\mathcal{B}.$\\
    		    Calculate the error $r+\gamma Q_{1} - Q_{2}$ and update $\hat{\vect{\theta}}_{2}.$\\}
    		 for every C steps toggle flag.\\
		}
		}
 \caption{coop-RL \label{alg1}}
\end{algorithm}

In this training strategy, each episode comprises in repetition of two phases. \emph{Phase 1: we first choose $NN_2$ to be target and $NN_1$ to be the actor.} For each play, the actor network is used to gather actions that are provided to the emulator. At the end of each play, the reward and the states are obtained from the emulator and $NN_1$ is updated using EDL update rule. Once, the first phase is completed, we initiate phase 2. \emph{Phase 2: we designate $NN_2$ to be the actor and $NN_1$ to be the target.} We alternatively switch between phase 1 and phase 2 for every $C$ plays and update the two networks. For each update, we sample a batch of data from the experience replay and use it to evaluate the error and update the weights. 

In the cooperative learning strategy, the two NNs are utilized to estimate the objective which is a function of TD error. Therefore, the asymptotic convergence of $J_E$~(Theorem~2) demonstrates the convergence of the TD error. To gain more insight into the convergence of the coop strategy, consider two cars~(Car 1 and 2) starting at two different points in a city. These cars do not know their destination but are able to communicate with each other. Furthermore, both cars are provided a reward value which describes the goodness of their location~((0)~no need to move, (1)~good direction, (-1)~bad direction). The goal of these cars is to minimize the TD error, which is defined as a function of target, actor and the reward value. Each car can either be the target or an actor.  When a car is target~(say Car 1), it broadcasts its location but does not move. On the other hand, the actor car~(say Car 2) exploits the targets' location and the reward value~(reinforcement for the actors' current location) to determine its movement. Car 2 seeks to move towards Car 1~(guided by location) while progressing towards the destination~(guided by the reward). After Car 2 has moved many times, it finds a new location, that is closer~(depending on the initialization) to the destination than the older location. The new and better location is used to guide Car 1 and therefore, Car 2 becomes the target and Car 1 becomes the actor. Using the reward and location, Car 1 can now move many times. Next, Car 1 becomes the target and Car 2 becomes the actor and the process continues. This process of learning reaches its culmination when the TD error is small, that is Car 1 and Car 2 have both reached a common location where the cost is zero. At this point, there is no incentive for moving~(cost~0) and cooperative strategy has converged.
\section{Implementation and Simulation Results}\label{sec: simulations}
\begin{figure}[h]
\includegraphics[width =0.9\columnwidth, keepaspectratio = True]{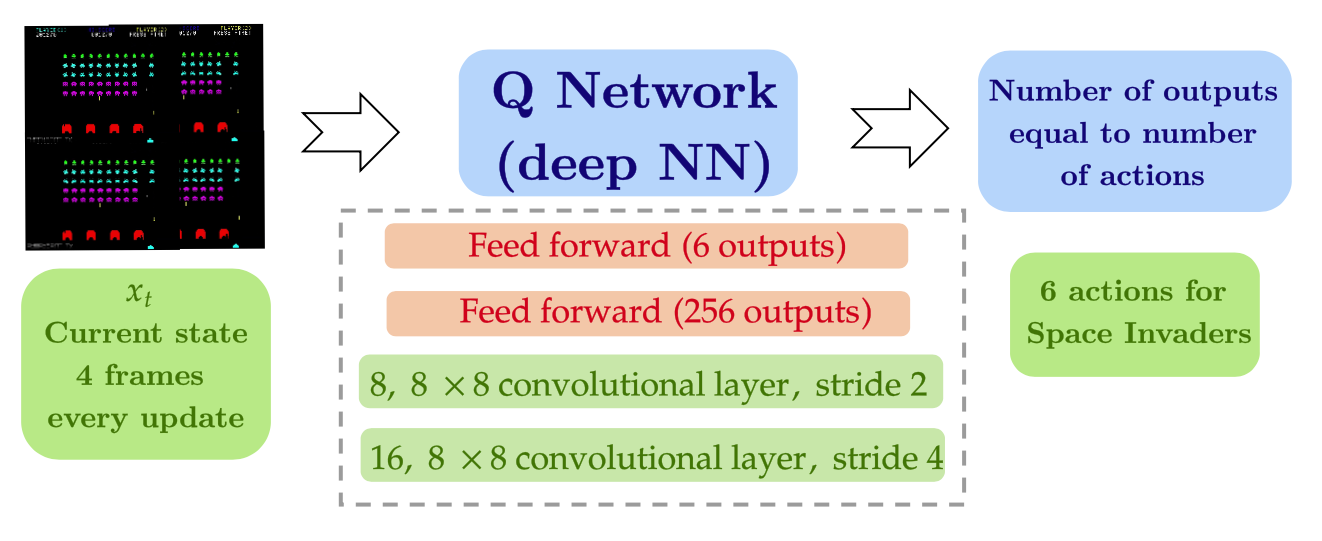}
    \caption{RL problem: Input-output flow diagram. }
    \label{fig:NN_Structure}
\end{figure}
To substantiate the efficacy of our approach, we consider the application of playing games. We choose a total of four games for our analysis, Cartpole, Breakout, Pong and Space Invaders. The game environments (emulators) for our analysis are provided by OpenAI and their details can be availed from the OpenAI website \cite{DBLP:journals/corr/BrockmanCPSSTZ16}. For all these games we consider two NNs with equal hyper-parameters. Our network architecture is depicted in Figure \ref{fig:NN_Structure}, where we use two convolutional layers and two feed-forward layers with \emph{relu} activation function. The input the network are images from the emulator whereas the output of the network are the Q-values corresponding to each action.

For comparisons, we consider three realizations. First, we consider the standard DQN with one NN~(target is a copy of the action network) and gradient driven updates, denoted as DQN. Second, we consider EDQL which is a NN~(target is a copy of the action network) with error driven updates \cite{Ragh2017error}.  The training strategy for these approaches is identical to the one proposed in \cite{mnih2015human}. Third, we consider the two network setup with standard gradient driven updates~(G-coop). For G-coop, we follow Algorithm 1, however, we use the Gradient-driven Adam optimizer instead of the error driven update rule.  Finally, we use the two network setup with error driven update and the coop strategy. As the two NNs are used to approximate function, from here on, we will refer to output of $NN_1$ as $Q_{1}$ and $NN_2$  as $Q_2.$ 

To record performance, we track the progression of the average reward~(instantaneous reward averaged over 100 episodes) and the cumulative rewards~(the average output of $Q_{1}$ and $Q_2$) with respect to the episodes. To record these values we do the following, at the end of each episode, we measure the performance of the network by recording the reward obtained by the network at the end of that episode. We also record the Q values for each action at the end of the episode. We report the mean and standard deviation of these quantities from the last 100 episodes. 

In these results on games presented below, the model is trained for a total of 100,000 episodes and each episode is composed of 1000 plays where each play provides a total of four images. Therefore, the NN has to process a total of 400 million frames. To provide insights into the inner workings of our approach, we choose the simpler, cartpole example as it is memory efficient. We start by discussing performance on the cart pole example and the NNs variants are implemented in Python 3.6 using  Pytorch libraries.  All simulations are performed using NVIDIA Tesla V100 SXM2 w/32GB HBM2 and NVIDIA Tesla K80 w/dual GPUs provided by the Argonne Leadership Computing Facility. 

\subsection{Example 1 -- Cartpole}
 One of most common examples in RL is that of cartpole~\cite{kumar2020balancing}. The problem is to prevent a vertical pole from falling by moving the cart left or right. Additional details can be found in the supplementary files. In our experiments, we execute our networks for a total of 10000 episodes. We record the $Q$ function values and the cumulative reward in each episode. 
 
\begin{figure}
  \centering
 \includegraphics[width = \columnwidth]{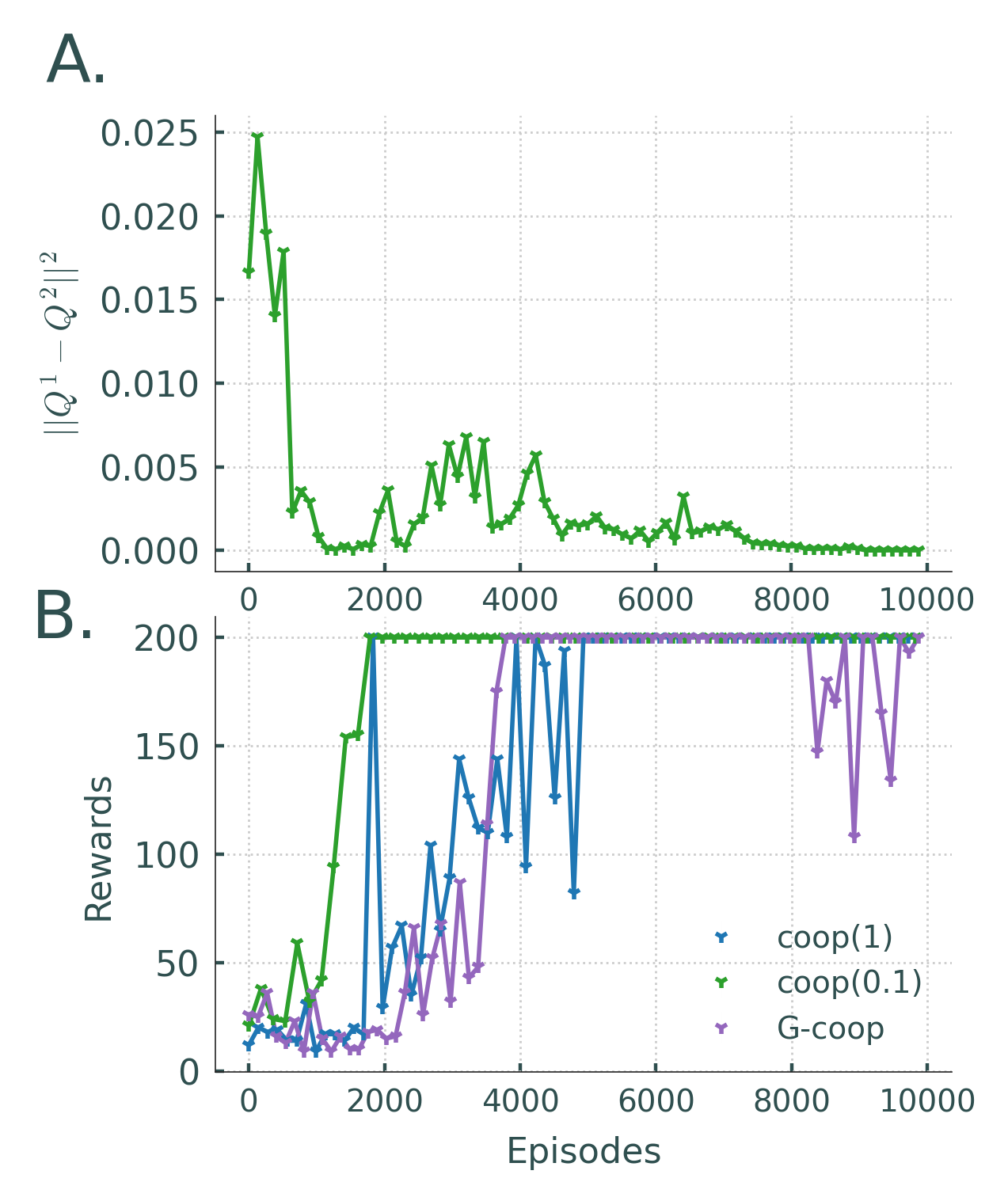}
\caption{Panel A: Mean rewards for different exploration rates, (exploration rate = 0) is the standard SGD, referred as $G-coop$, We use ADAM optimizer for this implementation.. Panel B:Difference between the Q-values generated by the two networks as a function of episodes.}
\label{fig:fig_balance}
\end{figure}

 We start by demonstrating that for our \textit{coop} approach the average reward over the last $100$ episodes is $200$~(which is the best reward for the cartpole problem) with a standard deviation of $0.2123,$ refer Table. \ref{tab:1}. For this experiment, we choose the hyper-parameters $ C = 30, len(\mathcal{B}) = 5000$ with an exploration rate of $s = 0.1$ that decays exponentially.  In Fig. \ref{fig:fig_balance}, panel A, we plot the progression of cumulative reward achieved by the coop method with the exploration rate of $0.1$, as a function of the total number of episodes. Cumulative rewards~( refer to coop(0.1)) steadily increase and reaches its peak value near $200$ around $2000$ episode. 
There are three crucial components that effect the performance of our methodology. First, we use two neural networks $Q_{1}, Q_2.$ Second,  the length of the experience replay buffer and third, the exploration rate.  We start by analyzing the behavior of the two neural networks, $Q_{1}$ and $Q_2.$.  Next, we will analyze, how these components affect the performance of our methodology. We start by analyzing the convergence of the two NNs
\subsubsection{Convergence of the two networks}
In our approach, we independently initialize the two networks and update them alternatively. Provided, an infinite number of iterations are provided, the $Q-$function values, predicted by the two networks are expected to converge to the optimal $Q$ function. To observe this behavior, we plot $\|Q_1 -Q_2\|$ as a function of the episodes in Fig. \ref{fig:fig_balance}. We point out that the normed difference between $Q_1$ and $Q_2$ values come to a value around $0.005$ around episode $2000$. We note that $2000$ episodes is where our approach reached the best cumulative reward values. 
\begin{figure*}
 \centering
 \includegraphics[width = \columnwidth,      keepaspectratio]{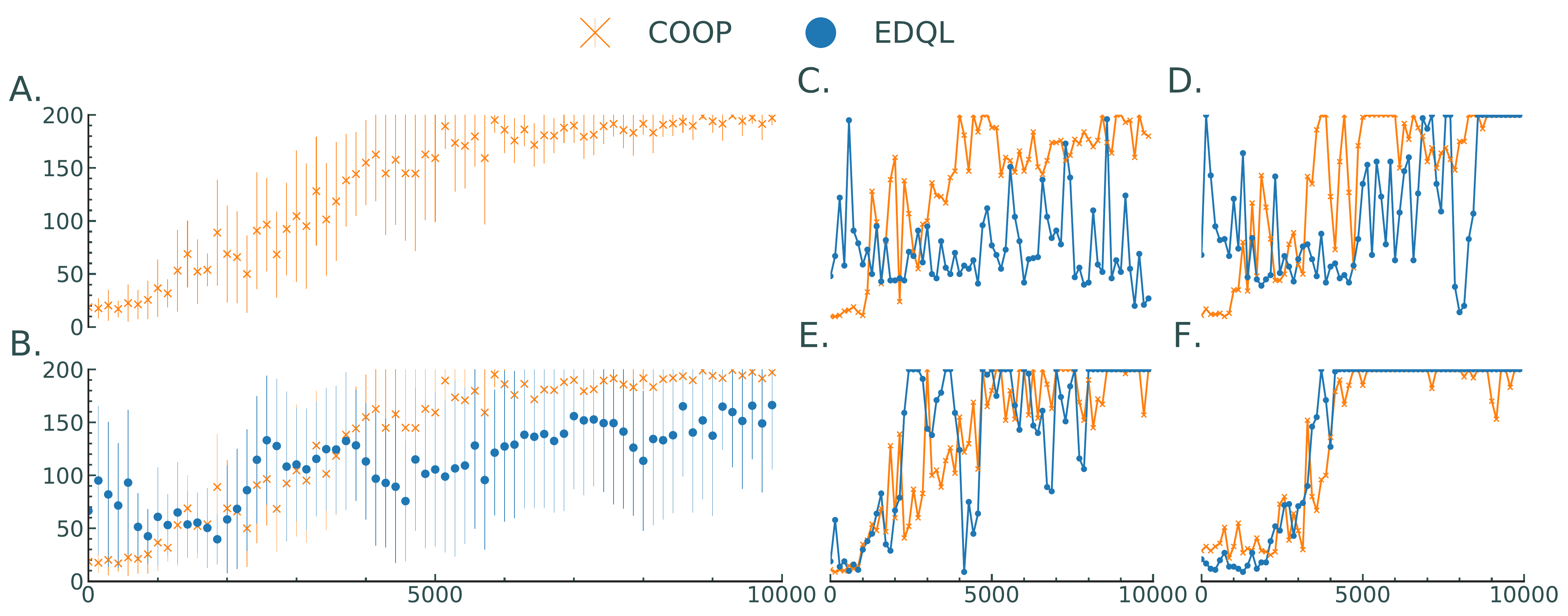}
  \caption{ Panel A-B: Mean Rewards (with error bars) depicting variance across different buffer sizes. Buffer sizes considered for this plot are 500,1000, 1500, 2000, 3000, 3500, 4000, 5000). As expected, the average rewards across different buffer sizes are better for the double NN-driven method instead of a single NN driven method. Panel C-F: Trend of rewards with respect to different buffer sizes.} 
  \label{fig:BufferSizes}
\end{figure*}
\subsubsection{Different Buffer Sizes}
Next, we analyze we compare the two NN approach~(coop) with one NN approach~(EDQL), both for various buffer sizes. In panels C-F, we plot the the trend of Rewards with respect to different episode for buffer sizes, 1000, 2000, 3000 and 4000. In Panel C-F, the markers indicate the mean of the cumulative reward over last 100 episodes.  Note that the best performance is observed for both EDQL and coop is obtained when the buffer size is 45000. 
 
However, we point that when the buffer size is reduced, the drop in performance for EDQL is much larger than coop. To observe this, note the trends from Panels F to C. We see a steady deterioration in performance. The plot shows that the use of the second NN allows to learn a representation for the data and thus provides convergence benefits. To further substantiate this, we observe Panel B. We note from Panel B that  the trend for coop~(Observe that this curve is plotted in Panel A as well) is consistently above the curve for the EDQL. The size of the buffer is a huge memory load in practical applications (both for read/write and storage). The use of a second neural network allows us to reduce this load by providing good performance for smaller buffer sizes.
 
\subsubsection{Effect of exploration}
Next, we analyze the effect of different exploration rate for the coop methodology and compare it to that of G-coop. First, we note that coop with exploration rate of 0 is equivalent to G-coop. The comparison is provided in Panel B of Fig. \ref{fig:fig_balance} where even a small exploration of $0.1$ improves the performance of  G-coop drastically. However, for a large exploration rate~($1$), coop converges quickly to 200 but oscillates significantly more in comparison to G-coop~(no exploration). On the other hand,  coop~(0.01) is stabler compared to coop~(0.5). These results concur with the common idea that larger exploration provides for a fast but unstable learning process. On the other hand, reasonable exploration can strike the right balance between exploration and exploitation. 
 
\subsection{RL Games}
For this part of analysis, we choose a total of three games: breakout, space invaders and pong, details can be found in \cite{mall2013self}. For comparisons, we choose DQN, EDQL, G-coop and coop. The hyper-parameters for this analysis is taken from \cite{mnih2015human} and kept consistent across different games. We record the mean and standard deviation values of the average reward in Table. \ref{tab:1}. Note that for all these games, the network architecture is identical to that proposed in DQN paper~\cite{mnih2015human}. Therefore, the input to the network are images.
 \begin{table*}[!t]
\centering
\begin{tabular}{p{2cm}p{2.4cm}p{3cm}p{3cm}p{3cm}}
\hline 
Method & Breakout & Space Invaders  & Cartpole &  Pong \\
\noalign{\smallskip}\hline\noalign{\smallskip}
DQN      &   \quad 287 (78)   & \quad 1139 (158)      &  \quad 193.6(0.71)    &  \quad 18.9(2.12)  \\
EDQL     &   \quad 290 (55)   & \quad 980  (163)      &  \quad 199.6(0.82)    &  \quad 19.11(1.88)  \\
G-coop &   \quad 314 (15)   & \quad 1174 (163)      &  \quad 200(0.82)      &  \quad 18.91(1.32)  \\ 
coop   &   \quad 313 (25)   & \quad 1144 (121)      &  \quad 200(0.2123)    &  \quad 19.15(2.1)  \\  
\noalign{\smallskip}\hline\noalign{\smallskip}
\end{tabular}
\caption{Mean score (standard deviation) over 100 episodes. \label{tab:1} }
\end{table*}

At the onset, we point that DQN is the baseline and provides reasonable cumulative rewards on all the games. The cumulative rewards obtained for DQN in our experiments are similar to the performance reported in \cite{mnih2015human}. Note from the table that we observe mean cumulative rewards with standard deviation $287~(78)$ for Breakout, $1139~(158)$ for space invaders, $193.6~(0.71)$ for cartpole and  $18.9~(2.12)$ for pong. A slight improvement is observed with EDQL everywhere except space invaders where we observe a $14$ \% drop in performance. The improvement is expected as EDQL introduces an inherent exploration strategy which provides convergence benefits.  However, one has to note that, even with an improved exploration strategy, best performance often depends on appropriate hyperparameter tuning which has not been done in this paper.

With the introduction of the second NN for learning the Q function, we observe a consistent improvement in performance indicated by better cumulative rewards in both G-coop and coop. For coop, we observe an improvement of  9 \% for breakout, 16 \% for space invaders, 3.3 \% for cartpole and 0.2 \% for pong. Similarly, we observe an improvement of  8 \% for breakout, 16 \% for space invaders, 0.2 \% for cartpole and 0.2 \% for pong.
\subsection{Discussions}
A common approach in the literature is the use of backpropagation/stochastic gradient descent~\cite{mnih2015human}. Typically, in such an approach, labels have to be defined. However, in RL-based design, the only feedback from the environment are the reward signals. To work with a definition of error, in most DRL schemes including ours, synthetic labels in the form of an optimal $Q$ function are defined as a function of the reward signal. Since the optimal $Q$ function is unknown and must be approximated using samples from the history. The error signal is a function of an approximated optimal $Q$ function, which provides imprecise information early in the learning phase.  If this imprecise  and small error is utilized as is done with stochastic gradient descent (SGD) method, the vanishing gradient problem will stagnate the learning. However, our approach guarantees non vanishing leaning signals. Furthermore, we introduce an exploratory strategy resulting in improved performance.

However, the improvement is contingent upon the idea that $Q^*$ is well approximated. The canonical way of obtaining this approximation is as follows: we start from the current value of Q~(that is the output of the RL agent). Since the optimal $Q$ value can only be greater than the current value in a maximization problem, we run multiple updates starting from the current value. These updates are essentially equivalent to running a Markov chain starting from the current value to obtain a new estimate of the $Q^*$ value.  This approximated $Q^*$ can be utilized as the target. Once new data is observed, one can repeat the procedure to get another approximation. To describe the error signal, we repeatedly replace the $Q^*$ with this running approximation of $Q^*$ through $Q$ which is essentially the output of the neural network. However, the main drawback of these canonical approaches is that, the process of approximating the optimal function is imperfect. To introduce enough exploration in the network for efficient exploration,  one needs to maintain a large experience replay buffer.

Although, to stabilize this approximation, approaches were introduced in \cite{dueling, doubleDQN}, they rely on the learning process of a single network that can still be biased by its own initialization. These constraints can be removed with the use of a second NN. If we run the same approximation procedure with two independent initialization, the solution is more robust as the optimal value is obtained only when the two networks converge to each other which is in contrast with \cite{dueling, doubleDQN}. This type of approach allows for the steady improvement in cumulative rewards we observed from Table. \ref{tab:1} results. Another crucial component of our methodology is the use of buffer and the use of the second network paired with the exploration strategy thus reducing the need for a large replay buffer which we demonstrated in Fig. \ref{fig:fig_balance}.

 An analysis of computational overhead, relative to the standard DQN approach, can be performed as follows, let $o[g]$ be the complexity of updating a neural network once using gradients. Since, each neural network is alternatively updated, at any point of time, only one network is being updated at a given time. Let the worst case complexity of evaluating the SVD of the gradient matrix be denoted by $o[S].$ Therefore, for every batch of data, the worst case complexity of our approach is $o[g+S]$. In comparison, the complexity of a standard DQN~\cite{mnih2015human} is $o[g].$


\section{Conclusion}\label{sec: conclusions}
In this paper, we present a two NN driven exploratory error driven learning approach for addressing DRL tasks. This approach presented the following advantages, (1.) we reduced the need for a large buffer sizes with the use of two NNs. (2) We demonstrated that the error driven strategy improves convergence by disentangling the learning across different layers in a deep NN. We provided both theoretical and experimental evidence to support this claim. (3) Our approach allows for methodical analysis of the learning problem and is extendable to diverse applications where a notion of error can be defined for an RL problem involving a dynamical process. 

Although significant improvements have been shown in this paper, there are major challenges that are still present. For instance, in this paper, we consider a typical reinforcement episodic learning setup where the RL agent attempts to learn from rewards due to both successes and failures. In real-time control applications such as controlling autonomous cars, failures are not acceptable as each failure can lead to enormous loss of revenue and worse, the loss of human life. 

 Despite the reduction in requirement of  buffer size, the size of experience replay buffer is still fixed at one million. Furthermore,  the larger the size of the experience array, the better the performance of the RL agent. There is a significant need of resources to achieve the required precision in learning optimal policies for a given task. Therefore, prior to deployment in real life control applications, the DRL methods must improve to the extent that learning is sufficiently precise with guaranteed convergence given limited time and resources.

\bibliographystyle{IEEEtran}
\bibliography{DFAWPE}

\end{document}